# *Amicus Plato, sed magis amica veritas*: There is a reproducibility crisis in COVID-19 Computational Fluid Dynamics studies

**Khalid M. Saqr, Ph.D.**

Innovation Hub, Arab Academy for Science, Technology and Maritime Transport, Al-Alamein Campus, 51718, EGYPT

Email: k.saqr@aast.edu

(ORCID: 0000-0002-3058-2705)

## ABSTRACT

There is overwhelming evidence on SARS-CoV-2 Airborne Transmission (AT) in the ongoing COVID-19 outbreak. It is extraordinarily difficult, however, to deduce a generalized framework to assess the relative airborne transmission risk with respect to other modes. This is due to the complex biophysics entailed in such phenomena. Since the SARS outbreak in 2002, Computational Fluid Dynamics (CFD) has been one of the main tools scientists used to investigate AT of respiratory viruses. Now, CFD simulations produce intuitive and physically plausible colour-coded results that help scientists understand SARS-CoV-2 airborne transmission patterns. In addition to validation requirements, for any CFD model to be of epistemic value to the scientific community; it must be reproducible. In 2020, more than 45 published studies investigated SARS-CoV-2 airborne transmission in different scenarios using CFD. Here, I systematically review the published CFD studies of COVID-19 and discuss their reproducibility criteria with respect to the CFD modeling process. Using a Weighted Scoring Model (WSM), I propose a novel reproducibility index for CFD simulations of SARS-CoV-2 AT. The proposed index ($0 \leq R_j^{CFD} \leq 1$) relies on three reproducibility criteria comprising 10 elements that represent the possibility of a CFD study ($j$) to be reproduced. Frustratingly, only 3 of 23 studies (13%) achieved full reproducibility index ($R_j^{CFD} \geq 0.9$) while the remaining 87% were found generally irreproducible($R_j^{CFD} < 0.9$). Without reproducible models, the scientific benefit of CFD simulations will remain hindered, fragmented and limited. In conclusion, I call the scientific community to apply more rigorous measures on reporting and publishing CFD simulations in COVID-19 research.

**Keywords:** COVID-19, SARS-CoV-2, airborne transmission, computational fluid dynamics, reproducibility

# INTRODUCTION

To the date of writing this article, COVID-19 pandemic outbreak has resulted in 1.6 million deaths and a historic crash of financial markets[1] leading to major fractures in the world economy[2]. This pandemic challenges our healthcare systems, economic models and social lifestyle[3]. Many countries refuged to enforcing nationwide curfew[4], travel ban and quarantine[5], social distancing and obligatory use of face masks[6] as measures to mitigate the outbreak. Despite few earlier controversies[7, 8], now there is a widely accepted theory among scientists now proposing that *airborne transmission (AT)* is a major infection scenario of COVID-19[9-11]. Few days after the WHO declared COVID-19 a pandemic[12], a study[13] published in New England Journal of Medicine demonstrated the possibility of SARS-CoV-2 AT experimentally. The study compared the stability of SARS-CoV-1 and SARS-CoV-2 in aerosols and on surfaces. It was evidently shown that SARS-CoV-2 has an aerosol stability similar to SARS-CoV-1 and remains infectious in aerosols for hours. SARS-CoV-2 infected patients were shown to exhibit high viral loads in the upper-respiratory tract[14], manifesting the possibility of producing highly infectious aerosols even from asymptomatic patients[15]. Infectious aerosols are often categorized according to the droplet particle size. During a sneeze or a cough, aerosols of respiratory tract fluid are produced containing large particles (i.e. droplets) typically greater than 5 μm in diameter. These particles impact directly on a susceptible individual. On the other hand, a susceptible individual could possibly inhale microscopic aerosol particles consisting of the residual solid components of evaporated respiratory droplets, which are small enough (<5 μm) to remain airborne for hours[16]. Even during speech, thousands of oral fluid droplets that constitute AT and COVID-19 infection risk[17]. It was also established that infectious SARS-CoV-2 RNA is persistent in aerosols collected in the vicinity of infected individuals with particles of small and large sizes[18]. Jin et al[19] showed that SARS-CoV-2 positive air samples can be collected in ICU room for four days after the residing patient tested negative. Guo et al[20] collected positive samples from ICU air as far as 4 m from patients. Despite rapid air changing in airborne infection isolation rooms (AIIRs), Chia et al[21] showed that SARS-CoV-2 RNA can be detected in air samples with particle sizes of $> 4\ \mu m$ and $1 - 4\ \mu m$. Razzini et al[22] confirmed the persistence of SARS-CoV-2 RNA in air samples taken from ICU room and corridor of a hospital in Milan, Italy. In highly populated communities and crowded spaces, AT could lead to catastrophic rise in infection probability[23, 24]. Therefore, investigating AT aerodynamics is of eminent importance to help mitigate infection risk at different scales and scenarios[25]. Computational Fluid Dynamics (CFD) is a very useful tool to manifest such importance by providing rapid evaluation method to identify AT risk at virtually any given scenario[26].

## RELEVANCE OF CFD MODELS IN SARS-2002 OUTBREAK

SARS-CoV-2 is an enveloped virus with a diameter of 0.1 $\mu m$, aerosol half-life of 1 hour and a concentration of $10^4 - 10^{11}$ RNAs/mL in respiratory fluids[27]. These are typical numbers for the Coronaviridae[28], including SARS-CoV-1 that caused the first pandemic of the 21$^{st}$ century. It has been established that viral RNA is carried in fluid particles produced by symptomatic patients while coughing and breathing[29] leading to AT of the virus. The degree to which AT constitutes infection risk depends on many variables that are yet to be comprehensively understood. The pathogenic similarity between the SARS-CoV-1 and SARS-CoV-2[30] establishes relevance between the role of AT in the two corresponding pandemics. The studies conducted on the 2002 SARS outbreak in Amoy Gardens housing complex of Hong Kong presented important insights. Yu et al[31] studied the distribution of the initial 187 cases of SARS in Hong Kong while searching for a possible transmission pathway to justify that cluster infection. Using logistic regression and CFD simulations, they showed that the infection risk pattern corresponded well with the predicted aerosol transfer pattern between apartments. In another study, Chu et al[32] analyzed the correlation between nasopharyngeal viral load of SARS patients on hospital admission and their geographic/location distribution in Amoy Gardens. They evidently concluded that AT played an important role in the Hong Kong outbreak. Li et al[33] used multi-zone airflow modeling to study the infection risk of 300 patients in the same housing complex. By comparing the virus concentration and aerosol dispersion predicted by their model and the distribution of infected patients in the housing blocks, they provided another evidence on AT. Moreover, Booth et al [34] conducted PCR tests on air samples collected from patient rooms during the Toronto outbreak. By detecting active SARS-CoV-1 in air around patients, they indicated AT possibility experimentally. These studies, all involving CFD simulations, showed that AT has played a significant role in SARS pandemic outbreak. This historic experience provided sufficient reasoning to consider AT risk of SARS-CoV-2 at the early beginning of COVID-19 outbreak.

# PURPOSE, SCOPE AND REVIEW APPROACH

The purpose of this article is to promote better reporting practice of CFD studies related to COVID-19 research and biomedical research at large. The scope is limited to the concept of reproducibility in CFD practice. The establishment of any CFD study requires proper level of verification, validation and reproducibility otherwise, it would not be possible to confirm the study's conclusion[35]. In COVID-19 air transmission research, CFD models represent the interface between the biomedical and managerial aspects of the COVID-19 infection control problem. The rapid, aggressive and mutating COVID-19 outbreak compels certain level of diligence to support balanced and effective decision making strategies of infection control. The scope of reproducibility here refers to the minimum level of details at which a CFD model could be reproduced with a quantifiable measure of error or deviation.

Similar to SARS-CoV-1, the importance of CFD in investigating SARS-CoV-2 AT is driven by its parametrization capability that identifies AT patterns of virtually any given scenario. Each scenario reveals aerodynamic aspects specific to the settings, physical models and validation data of the CFD simulation. To compile a comprehensive resource of COVID-19 CFD studies and their reproducibility criteria, a systematic review of the CFD studies is conducted and discussed. The search engine **Scopus** was used to identify the reviewed studies. On December 23$^{rd}$ 2020, a search in *TIT-ABS-KEY* field yielded 49 studies with the search string: TITLE-ABS-KEY ( "computational fluid dynamics" OR "CFD" ) AND TITLE-ABS-KEY ( "Covid-19" OR "SARS-CoV-2" OR "Coronavirus" ) AND ( LIMIT-TO ( PUBYEAR , 2021 ) OR LIMIT-TO ( PUBYEAR , 2020 ) OR LIMIT-TO ( PUBYEAR , 2019 ) ). The studies were obtained, reviewed to identify the main corpus of this review. A total of 23 studies were found to report formal and conclusive CFD results. The studies were then classified according to transmission scenario, solver, computational model, physics, and settings of each respective CFD model. The studies were classified into three sets. Set (A)[36-43] represents the studies conducted on COVID-19 in healthcare facilities such as hospital wards and care rooms. Set (B)[44-52] represents the studies conducted on SARS-CoV-2 AT in respiratory scenarios such as in nasopharyngeal and pulmonary spaces. Finally, set (C)[53-58] represents the studies conducted on COVID-19 in generic spaces and buildings.

**Bibliometric insight**

By subject area, 24% of the studies were published in engineering journals while the publications in medicine, chemical and environmental engineering were 16%, 12% and 11%, respectively. It must be noted that many studies are published under more than one subject area. The 49 studies received 117 citations in total achieving a Scopus h-index value of 5 until the December 23rd. The studies published under medicine research area received 52 citations (44% of the total citation count). Eight of the 23 CFD studies received 86 citations (73.5% of the total citation count). By article type, the 49 studies were classified into 80% articles, 6% reviews and 14% of other article types. Overall, self-citations constituted only 8.5% of the total citation count. Figure 1 shows a graphical representation of bibliometric data. The supplementary materials include tabulated data of the citation count of the 49 studies.

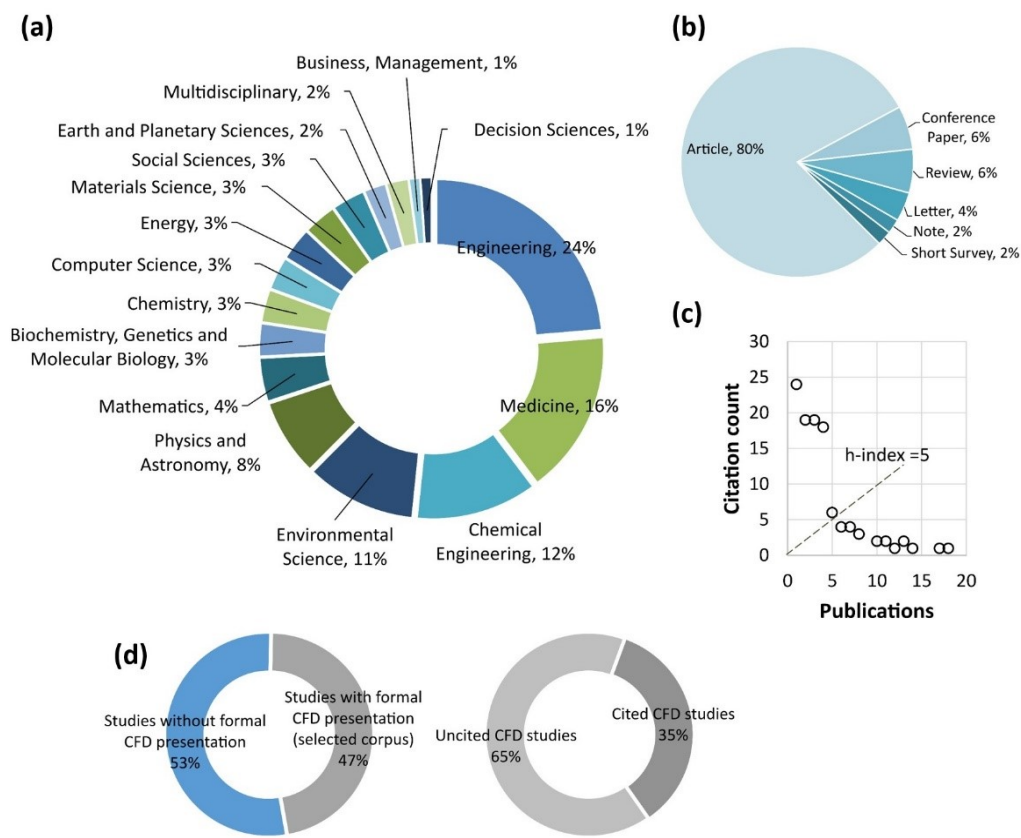

**Figure 1.** Bibliometric insight into the COVID-19 CFD published studies showing (a) contributions by subject area, (b) by publication type and (c) h-graph indicating h-index value of 5 and (d) percentage of cited CFD studies compared to the selected corpus.

# METHOD: A REPRODUCIBILITY INDEX FOR CFD MODELS OF COVID-19

There are three criteria of reproducibility that any fluid dynamicist with firsthand experience in modern CFD software needs to address in order to replicate a simulation case[59, 60]. The first criterion deals with the numerical formulation of the model, the second one deals with physical formulation and the third deals with the parametric framework of the simulation being reproduced. The elements of each reproducibility level are detailed in table 1. In order to evaluate the 23 CFD studies, a novel reproducibility index is proposed. The index is developed based on the weighted scoring model (WSM) commonly used in engineering project management. The importance of each element relevant to COVID-19 research is assigned a numerical weight $w = [5, 10, 20]$ representing low, medium and high importance to the reproducibility process, respectively. Each element in each study is assigned a binary score $s = [0,1]$ to represent the reproducibility of each element. Then, the CFD reproducibility index is calculated as:

$$R_j^{CFD} = \frac{\sum_{i=1}^{n} w_i s_i}{\sum_{i=1}^{n} w_i} \qquad (1)$$

where $j$ is the study index and $n$ is the number of reproducibility elements, respectively. Equation (1) dictates that $0 \leq R_j^{CFD} \leq 1$ where values at 0 and 1 represent entirely irreproducible and reproducible simulations, respectively. The elements relevant to COVID-19 research and the scope of this study and their corresponding weights are detailed in table (2). The reasoning proposed for the weights of reproducibility elements is intuitive and derived from the CFD modeling process. Two elements of high weight ($w_i = 20$) represent the mandatory information required for the setup of the CFD simulation which are the model dimensions and initial and boundary conditions (40% of the total weight per study). Five elements with medium weight ($w_i = 10$) represent the information that would make reproducibility possible (50% of the total weight per study). Two elements of low weight ($w_i = 5$) represent the information that would make the reproducibility process easy (10% of the total weight per study). Here, we argue that for a study to be reproducible $R_j^{CFD} \geq 0.9$. Irreproducible studies have $R_j^{CFD} \leq 0.4$ while studies with $0.4 < R_j^{CFD} < 0.9$ are difficult to reproduce as they lack information necessary to perform CFD simulation.

**Table 1. Levels and elements of CFD simulation reproducibility process**

| Reproducibility Level | Elements |
|---|---|
| 1. Numerical formulation | 1.1. Solver and/or iterative formulation <br> 1.2. Coupling and discretization of the governing equations <br> 1.3. Computational domain and grid (i.e. mesh) <br> 1.4. Time stepping <br> 1.5. Solution convergence criteria |
| 2. Physical formulation | 2.1. Physical models (turbulence, energy, heat transfer, multiphase flow, magnetism…etc) <br> 2.2. Physical properties of working fluids <br> 2.3. Initial and boundary conditions |
| 3. Parametric framework | 3.1. Independent variables and their values <br> 3.2. Dependent variables and their representation <br> 3.3. Validation of the CFD model and results |

**Table 2. COVID-19 CFD reproducibility elements and their corresponding WSM weights**

| | Element | $w_i$ | $S_i$ |
|---|---|---|---|
| **Numerical formulation** | Solver | 10 | 1 if the solver was reported <br> 0 if the solver was not reported |
| | Model Dimensions | 20 | 1 if the dimensions were reported <br> 0 if the dimensions were not reported |
| | Grid resolution | 10 | 1 if the grid resolution was reported <br> 0 if the grid resolution was not reported |
| **Physical formulation** | Turbulence model | 10 | 1 if the turbulence model was reported <br> 0 if the turbulence model was reported |
| | Aerosol model | 10 | 1 if the aerosol model was reported <br> 0 if the aerosol model was not reported |
| | Reynolds number | 5 | 1 if the Reynolds number was reported <br> 0 if the Reynolds number was not reported |
| | Initial and Boundary conditions | 20 | 1 if the initial and boundary conditions are FULLY reported <br> 0 if any initial and/or boundary conditions are missing |
| | Particle density | 5 | 1 if the particle density was reported <br> 0 if the particle density was not reported |
| **Independent variables** | Aerosol particle diameter | 5 | 1 if the particle diameter is reported <br> 0 if the particle dimeter is not reported |
| | Validation | 5 | 1 if the study reports validation <br> 0 if the study does not report validation |
| | $\sum_{i=1}^{n} w_i = 100$ | | $0 \leq \sum_{i=1}^{n} w_i S_i \leq 100$ |

# RESULTS AND DISCUSSION

The distribution of $R_j^{CFD}$ among the 23 studies comprising sets A, B and C are graphically presented in figure 2. The detailed calculations and scoring sheet is provided in the supplementary materials. The mean value of $R_j^{CFD}$ was 0.62, 0.57 and 0.71 for sets A, B and C, respectively. The values of $R_j^{CFD}$ of the 23 studies were found to be normally distributed around mean (mean=0.62) as shown in figure 3. The distribution, however, is slightly skewed for $R_j^{CFD} < 0.62$. Three studies, one from each set[42, 46, 57], achieved full reproducibility score $\left(R_j^{CFD} \geq 0.9\right)$ while six studies[40, 41, 44, 47, 52, 53] were found to be irreproducible $\left(R_j^{CFD} \leq 0.4\right)$. The remaining 14 studies were found to have a reproducibility score in the range of $0.4 < R_j^{CFD} < 0.9$. These studies are difficult to reproduce as they lack important information. The difficulty of reproducing these studies varies according to the missing information. Figure 4 shows the missing and available information in the 14 studies according to the elements presented in table 2. Missing information about Reynolds number and aerosol particle diameter characterize 79% and 36% of these studies, respectively. Missing dimensions and validation information characterize 50% of such studies while the remaining reproducibility elements vary from one study to the other.

Infection control in healthcare facilities is of crucial importance in managing COVID-19 outbreak. Set (A) [36-43] comprises CFD studies of different AT scenarios in hospitals and healthcare facilities. The computational domain in these studies always represent the air flow around aerosol source of particular settings that represent SARS-CoV-2 AT. Grid resolution ranged from $0.9 \times 10^6$ in simple two-dimensional representation of generic care room[38] to $50 \times 10^6$ cells in three-dimensional representations in prefabricated inpatient ward[42]. The use of RANS models was the main approach to model turbulence with just two studies were reported using LES[38, 39]. Aerosol modeling was predominantly conducted using the Eulerian-Lagrangian approach[61]. Only one study reported the Reynolds number value[38] and another reported aerosol particle density[36] of the simulation. Four studies[36, 41-43] reported the aerosol particle size in the simulations with a range from 0.69 to $500 \mu m$. Relatively similar approaches were identified in the studies comprising set (C)[53-58] where SARS-CoV-2 AT was studied in generic building spaces. On the other hand, the respiratory studies comprised in set (B)[44-52] had some different CFD approaches. The computational domain in such studies represented either expiratory space[44, 46, 50] (i.e. space around person's head) or inspiratory space[45, 51] (i.e. tracheobronchial or nasopharyngeal space). Grid resolution, therefore, had a wide range from $0.15 \times 10^6$ to $650 \times 10^6$ cells[46, 49]. The aerosol particle diameter in set (C) studies[44, 45] ranged from 0.1 to 525 $\mu m$. Turbulence and aerosol models were dominated by RANS and Eulerian-Lagrangian approaches similarly with sets (A) and (C). These CFD studies should empower our understanding of COVID-19 outbreak[7, 26, 62, 63]. This is demonstrated by the citation data of these studies, as presented earlier in this article. However, and unlike other important testing and characterization tools addressing this unprecedented pandemic, published CFD studies suffer from a lack of sufficient reproducibility criteria to advance infectious aerosol research. With more than half of the studies missing information about domain dimensions, Reynolds number, particle density and validation; a reproducibility crisis is identified and should be addressed. It is noteworthy to mention that none of the studies included in this review has provided any form or format of digital files to enable manipulation and processing of the CFD results.

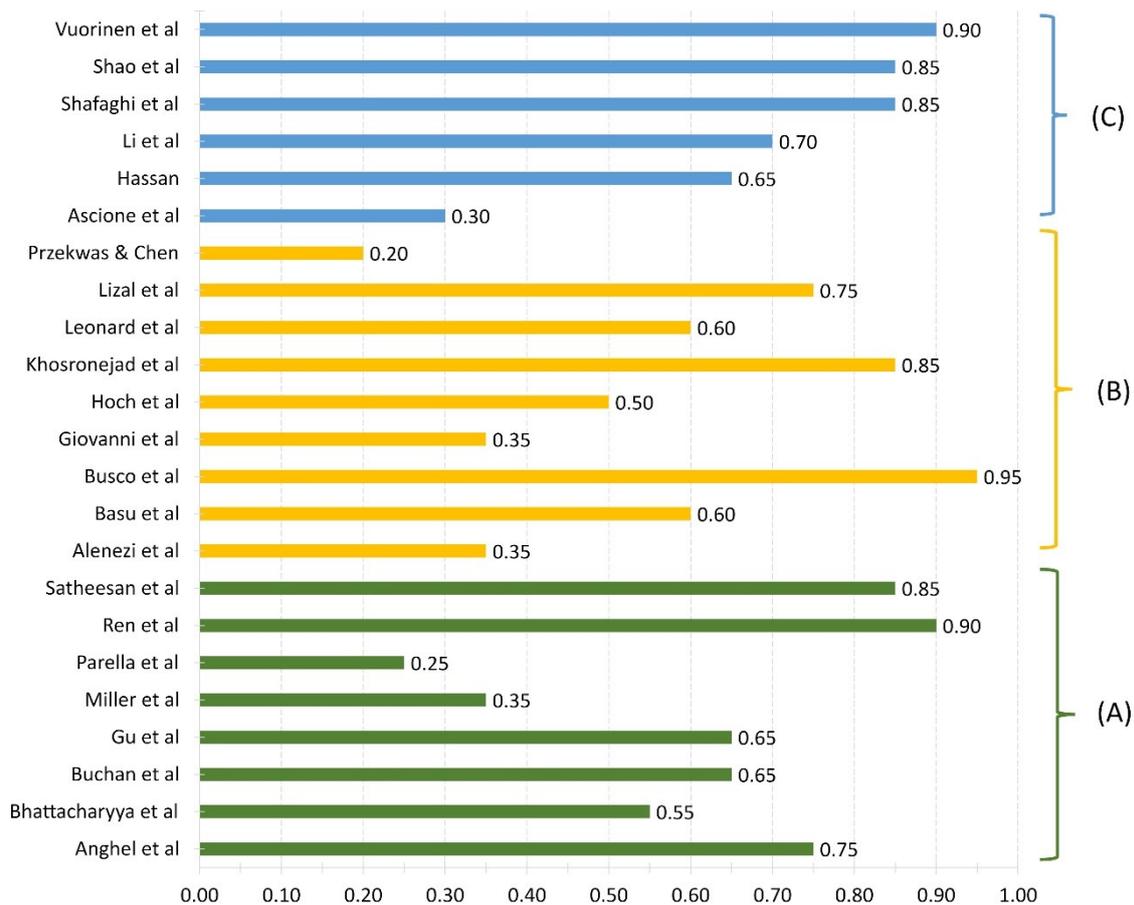

**Figure 2**. Reproducibility index $\left(R_j^{CFD}\right)$ of the 23 CFD studies (sets A, B and C) of COVID-19 airborne transmission.

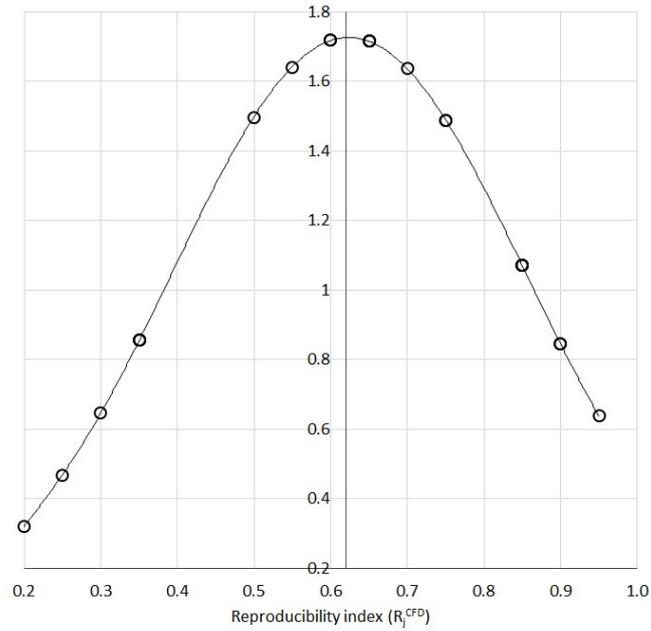

**Figure 3.** Normal distribution of $R_j^{CFD}$ values of the 23 studies (mean = 0.62, standard deviation = 0.2, median = 0.65)\

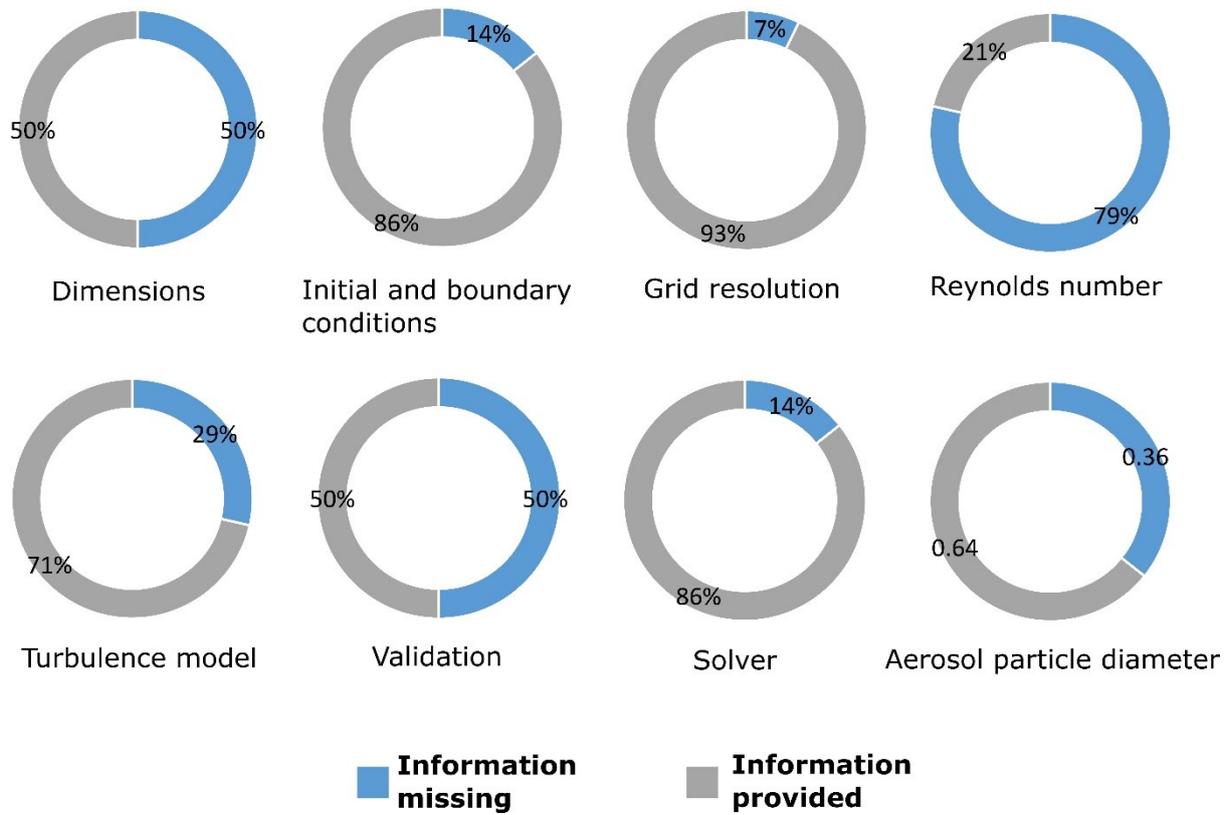

**Figure 4.** Available and missing information of the reproducibility elements in 14 studies that have reproducibility index in the range of $0.4 < R_j^{CFD} < 0.9$.

# CONCLUSION

A novel reproducibility index $\left(R_j^{CFD}\right)$ is proposed to measure the possibility of a CFD study to be reproduced. The index accounts for three criteria comprising a total of 10 elements that comprise the essential information required for reproducing CFD simulation. When the index was applied to 23 published CFD studies related to COVID-19, it revealed a reproducibility crisis. The results were normally distributed around a mean value of 0.62 and revealed that only 13% of the selected studies achieve the reproducibility criteria ($R_j^{CFD} \geq 0.9$). In conclusion, we propose this novel reproducibility index as a criteria for publishing CFD studies related to COVID-19 research in order to empower reproducibility and validation in this important research topic.

**CONFLICT OF INTEREST:** NONE

**FUNDING:** NONE